\documentstyle[11pt,aaspp4]{article}

\slugcomment{Accepted by the Astrophysical Journal Letters on 12/11/97}

\lefthead{Mighell, Sarajedini, \& French}
\righthead{SMC Intermediate-Age Populous Cluster NGC 416}

\begin{document}

\def\et{\hbox{et~al.\ }}
\def\vs{\hbox{vs }}
\def\feh{\hbox{[Fe/H]}}
\def\bmv{\hbox{$B\!-\!V$}}
\def\ebmv{\hbox{E$(\bmv)$}}
\def\dbmv{\hbox{$d_{(B\!-\!V)}$}}
\def\dbmr{\hbox{$d_{(B\!-\!R)}$}}
\def\mvto{\hbox{$M_V^{\rm{TO}}$}}
\def\mvrhb{\hbox{$M_V^{\rm{RHB}}$}}
\def\vhb{\hbox{$V_{\rm{HB}}$}}
\def\lea{\mathrel{<\kern-1.0em\lower0.9ex\hbox{$\sim$}}}
\def\gea{\mathrel{>\kern-1.0em\lower0.9ex\hbox{$\sim$}}}
%
\def\ifundefined#1{\expandafter\ifx\csname#1\endcsname\relax}

\title{WFPC2 OBSERVATIONS OF THE SMALL MAGELLANIC CLOUD
INTERMEDIATE-AGE POPULOUS CLUSTER NGC 416\altaffilmark{1}
}

\author{
{\sc
Kenneth J. Mighell\altaffilmark{2},
Ata Sarajedini\altaffilmark{3,}\altaffilmark{6},
and
Rica S. French\altaffilmark{4}
}
}
\affil{Kitt Peak National Observatory,
National Optical Astronomy Observatories\altaffilmark{5},\\
 P. O. Box 26732, Tucson, AZ  85726 \\ {\it mighell@noao.edu, ata@noao.edu,
rfrench@noao.edu}}

\altaffiltext{1}{
Based on observations made with the NASA/ESA
{\em{Hubble Space Telescope}},
obtained from the data archive at the Space Telescope
Science Institute,
which is operated by the Association of
Universities for Research in Astronomy, Inc.\ under NASA
contract NAS5-26555.
}

\altaffiltext{2}{
Guest User, Canadian Astronomy Data Centre, which is operated by the
Dominion Astrophysical Observatory for the National Research Council of
Canada's Herzberg Institute of Astrophysics.
}

\altaffiltext{3}{Hubble Fellow}

\altaffiltext{4}{Based on research conducted at NOAO as part of the
Research Experiences for Undergraduates program.}

\altaffiltext{5}{NOAO is operated
by the Association of Universities for Research in Astronomy, Inc., under
cooperative agreement with the National Science Foundation.}

\altaffiltext{6}{Current address:
Department of Physics and Astronomy,
San Francisco State University,
1600 Holloway Avenue,
San Francisco, CA~~94132.
Electronic mail: 
ata@stars.sfsu.edu
}

\clearpage
\begin{abstract}
We present our analysis of archival
{\sl{Hubble Space Telescope}} Wide Field Planetary Camera 2
observations
in
F555W ($\sim$$V$)
and
F450W ($\sim$$B$)
of the intermediate-age populous star cluster NGC 416 in the
Small Magellanic Cloud galaxy.
We use published photometry of two other SMC populous star clusters,
Lindsay 1 and Lindsay 113, to investigate
the age sequence of these three star clusters.
We estimate
that these clusters have age ratios of
$age_{\rm{NGC416}}/age_{\rm{L1}} \approx 0.73\!\pm\!0.05$
and
$age_{\rm{L113}}/age_{\rm{L1}} \approx 0.52\!\pm\!0.09\,$
using an extrapolation of the $\dbmv$ method
(which uses the color difference between the red horizontal branch
and the red giant branch as an age indicator)
of Sarajedini, Lee, \& Lee [ApJ, 450, 712 (1995)]\,.
These age ratios provide absolute age estimates of
6.6$\pm$0.5 Gyr
and
4.7$\pm$0.8 Gyr
for
NGC 416
and
Lindsay 113,
respectively,
assuming that Lindsay 1 is 9 Gyr old.
Metallicities of
$\feh\!=\!-1.44$$\pm$$0.12$,
$-1.35$$\pm$$0.08$,
$-1.24$$\pm$$0.11$ dex,
and reddenings of
$\ebmv\!=\!0.08$$\pm$$0.03$,
$0.06$$\pm$$0.02$,
$0.00$$\pm$$0.02$ mag
for NGC 416, Lindsay 1, and Lindsay 113, respectively,
were determined using the simultaneous reddening and metallicity (SRM)
method of Sarajedini \& Layden [AJ, 113, 264, (1997)].
Accurate (relative) ages
for the intermediate-age populous
clusters in the Small Magellanic Cloud
(e.g.\ via deep main sequence photometry)
would allow the $\dbmv$ method
to be recalibrated with
star clusters that are significantly younger than 7 Gyr.
An
extended $\dbmv$ method could
prove to be a very useful age diagnostic for future studies of
the intermediate-age metal-poor stellar populations in
Local Group galaxies where accurate main-sequence turnoff photometry at
$M_V\!\approx\!+4$ mag is currently not possible or practical.
\end{abstract}

\keywords{
    galaxies: clusters: individual (Lindsay 1, Lindsay 113, NGC 416)
--- galaxies: individual (Small Magellanic Cloud)
--- galaxies: star clusters
--- stars: Population II
}

\clearpage
\section{INTRODUCTION}

The Small Magellanic Cloud (SMC) has at least 6 populous star
clusters with $\feh < -1.0$:
NGC 121, NGC 416, Kron 3, NGC 339, Lindsay 1, Lindsay 113.
To within the measurement errors, they all have metallicities in the
range $-1.4 < \feh < -1.1$ and horizontal branch (HB) morphologies that are
predominantly redward of the RR Lyrae instability strip.
Ground-based observations suggest that these clusters have significantly
different ages despite their similar metallicities.
For example, Stryker \et (\cite{stet1985})
estimate that NGC 121 is $\sim$12 Gyr old.
In the case of NGC 339, Mould \& Aaronson  (\cite{moaa1980})
determined an age between
3 and 6 Gyr while Elson \& Fall (\cite{elfa1985})
find that it is $\sim$11 Gyr old.
The ages of the clusters Lindsay 113, Lindsay 1, and Kron 3
fall in the range between $\sim$6 and $\sim$12 Gyr
(Olszewski \et \cite{olet1996}).
Durand \et (\cite{duet1984})
estimated that the age of NGC 416 is 2.5$\pm$0.7 Gyr;
one year later
Elson \& Fall (\cite{elfa1985})
estimate an age of 5.9$\pm$1.9 Gyr
while Bica \et (\cite{biet1985}) estimate that it is 8$\pm$1.5 Gyr old.
The uncertainty in the ages of these systems underscores just how
little we know about the early evolutionary history of the SMC.

In this paper, we report on the first CCD observations of the SMC
populous star cluster NGC 416.
We estimate and compare various properties of
NGC 416, Lindsay 1, and Lindsay 113,
in order to improve our knowledge of the
the formation history
of the Small Magellanic Cloud.

\section{OBSERVATIONS AND PHOTOMETRIC REDUCTIONS}

The SMC globular cluster NGC 416
was observed with the
{\sl{Hubble Space Telescope}} Wide Field Planetary Camera 2 (WFPC2)
on 1994 February 6 through the
F450W ($\sim$$B$)
and F555W ($\sim$$V$)
filters.
The WFPC2 PC1 aperture
(Biretta \et \cite{biet1996})
was centered on the target
position of
$\alpha = 01^{\rm h}\ 07^{\rm m}\ 59^{\rm s}$
and
$\delta = -72\arcdeg\ 21\arcmin\ 26\arcsec$
(J2000.0)
and
two high-gain observations were obtained:
400-s in F450W
and
200-s in F555W.
The two datasets
(F450W: U26M0501T;
F555W: U26M0502T)
were recalibrated at the Canadian Astronomy Data Centre
and retrieved electronically by us using a guest account which was
kindly established for KJM.
The data were analyzed with
the CCDCAP digital circular aperture
photometry code developed by Mighell to analyze
{\sl{HST}}
WFPC2
observations (Mighell \cite{mi1997}, and references therein).
The instrumental magnitudes were transformed to Johnson $B$ and $V$ using
the standard WFPC2 equations given
in Tables 10 and 7, respectively, of Holtzman \et (\cite{hoet1995b}).
These color equations were recently used by
Mighell \et (\cite{miet1996})
to show that the age of Large Magellanic Cloud cluster
Hodge 11 is identical to that of the Galactic globular cluster M92
with a relative age-difference uncertainty ranging from 10\% to 21\%.
Complete details of the photometric reduction of these NGC 416 observations
are given in Mighell \et (\cite{miet1998}).

In the following sections,
we compare the cluster properties of
NGC 416 with two other SMC populous clusters, Lindsay 1
and Lindsay 113.
Before we can proceed with this comparison,
we require $BV$ photometry for all three clusters.
While $BV$ photometry exists for Lindsay 1 (Olszewski \et \cite{olet1987}),
none is available in the literature for Lindsay 113.
We converted the $BR$ photometry of Mould \et (\cite{moet1984})
to the $BV$ system using the color transformation equation\footnote{
By analyzing bright stars from many globular clusters over a wide
range of metallicities spanning a color range
$-0.2 \leq (\!B\!-\!R) \leq 2.5$ mag,
Sarajedini \& Geisler (\cite{sage1996})
found the following color transformation equation,
$
(B\!-\!V)_o
=
0.01135
+ 0.6184(\!B\!-\!R)_o
- 0.1071(\!B\!-\!R)_o^2
+ 0.1249(\!B\!-\!R)_o^3
- 0.0319(\!B\!-\!R)_o^4
$,
which has an rms error of 0.018 mag.
}
of Sarajedini \& Geisler (\cite{sage1996}).

\section{COLOR-MAGNITUDE DIAGRAMS}

The $V$ \vs $\bmv$ color-magnitude diagrams (CMD) of our observed stellar
field in NGC 416 reach $V\!\approx\!24$ mag and are displayed in
Fig.\ \ref{fig-1}.
We have arbitrarily split our observations into two regions:
(1) the ``cluster'' region
[PC1 CCD: see Fig.\ \ref{fig-1}b]
and
(2) the SMC ``field'' region
[WF2, WF3 and WF4 CCDs: see Fig.\ \ref{fig-1}c].
We statistically removed the SMC field population from the cluster region
CMD (Fig.\ \ref{fig-1}b) following the procedure
described in Mighell \et (\cite{miet1998})
which is
similar to, but more conservative than,
that used in the Hodge 11 study of Mighell \et (\cite{miet1996}).
A total of 2826 stars are
probable cluster members and they are displayed in the cleaned
cluster CMD (see Fig.\ \ref{fig-1}d).
This CMD cleaning method is probabilistic and
{Fig.\ \ref{fig-1}d} therefore
represents only one out of an infinite number of different possible
realizations of the cleaned NGC 416 CMD.

\section{REDDENINGS AND METALLICITIES}

The metallicity and reddening of a globular cluster can be determined
simultaneously, in an internally consistent manner,
using the magnitude level of the horizontal branch (HB),
the color of the red giant branch (RGB) at the level of the HB,
along with the shape and the position of the RGB
(Sarajedini \& Norris \cite{sano1994}).
The simultaneous reddening and metallicity (SRM)
method has subsequently been developed for use with several standard
filter combinations:
$V$ \vs $V\!-\!I$ (Sarajedini \cite{sa1994}),
$R$ \vs $B\!-\!R$ (Sarajedini \& Geisler \cite{sage1996}),
and
$V$ \vs $B\!-\!V$ (Sarajedini \& Layden \cite{sala1997}).
We now apply the latter version to our photometry of NGC 416.
First, we estimate the magnitude of the NGC 416 red HB by
considering all stars with
$19.5\!<\!V\!\!<\!20.0$
and
$0.70\!<\!B\!-\!V\!\!<\!0.84$ mag.
This yields $\vhb = 19.74$$\pm$$0.05$ mag, where
the quoted error reflects our estimate of the uncertainty in the
photometric zeropoint. We find that the mean color of the HB
is $(\bmv)_{\rm{HB}} = 0.786$$\pm$$0.003$ mag; the error
being the standard error of the mean.
We fitted a 4th order polynomial to the RGB stars,
using the iterative 2$\sigma$ rejection
algorithm described by Sarajedini \& Norris (\cite{sano1994}),
which gives the color of the RGB at the level of the HB
as $(\bmv)_g = 0.913$$\pm$$0.034$ mag.
Using the SRM method with these parameters,
we determine that the metallicity and reddening of NGC 416 are
$\feh = -1.48$$\pm$$0.19$ dex
and
$\ebmv = 0.09$$\pm$$0.06$ mag, respectively, where the errors have been
estimated following the prescription of Sarajedini (\cite{sa1994}).

The metallicity of a cluster can also be determined from the
slope of the RGB. It has been well established
that more metal-rich clusters possess RGBs with shallower slopes.
To quantify this behavior, we follow the work of Hartwick (\cite{ha1968})
and define the RGB slope with the parameter
$S_{-2.5} \equiv 2.5 / \left[\,(\bmv)_{-2.5} - (\bmv)_g \,\right]$,
where $(\bmv)_{-2.5}$ designates the $\bmv$ color of the RGB at a point 2.5
magnitudes brighter than the HB.
We find the relation between metallicity, $\feh$,
and the globular cluster RGB slope parameter, $S_{-2.5}$, to be
$\feh = 0.569 - 0.419S_{-2.5}$,
from a weighted least-squares fit to the
primary and secondary calibrating globular clusters
(excluding Lindsay 1) of Sarajedini \& Layden (\cite{sala1997}).
The root-mean-square deviation between
this relation and the data is $\sigma_{\rm{[Fe/H]}} = 0.15$ dex
(Mighell \et \cite{miet1998}).
Utilizing the previously derived polynomial fit to the RGB,
we find that $(\bmv)_{-2.5} = 1.440$$\pm$$0.020$ mag and
$S_{-2.5} = 4.74$$\pm$$0.36$ for NGC 416, and
therefore calculate a metallicity of
$\feh = -1.42$$\pm$$0.15$ which is in good agreement with the results
of the SRM method.
A reddening estimate of E$(\bmv) = 0.08 \pm 0.04$ is then derived using
Eq.\ (1) of Sarajedini \& Layden (\cite{sala1997}).
A weighted mean of the SRM method results and that derived via the RGB
slope analysis gives
$\feh = -1.44$$\pm$$0.12$ and $\ebmv = 0.08$$\pm$$0.03$ mag,
which will serve as our adopted values for NGC 416 in the remainder of
this paper.

Following the above procedures,
we applied the SRM method and the RGB slope method to the $BV$ photometry of
Lindsay 1 and Lindsay 113
in order to estimate their metallicities and reddenings
(see Table 1).
The fact that our metallicity and reddening estimates for
Lindsay 1 and Lindsay 113 are fully consistent with the values
previously adopted by the authors of that photometry
suggests that our metallicity and reddening analysis procedures
provide reasonable and believable results.

\section{AGES}

The most robust age determination techniques are those that deal with the
measurement of relative ages. As such, we have chosen to study the
age of NGC 416 relative to two other SMC populous clusters, Lindsay 1
and Lindsay 113.

In the previous section, we determined that
NGC 416, Lindsay 1, Lindsay 113 have similar metallicities
($\feh \approx -1.35$ dex).  If we make the naive assumption
that these clusters are also the same age, then their horizontal
branches should have the same luminosity.
Likewise,
if the $BV$ cluster photometry for NGC 416 and Lindsay 113 were shifted
to give the same $V$ magnitude for the horizontal branch, $V_{\rm{HB}}$,
then
the apparent $V$ turnoff magnitudes, $V_{\rm{TO}}$,
of these clusters would also be identical.
Figure \ref{fig-2} performs this experiment.

Our $V$ \vs $\bmv$ fiducial sequence of
Lindsay 1 is compared with our transformed $BV$
photometry of Lindsay 113 in Fig.\ \ref{fig-2}a.
Similarly,
our Lindsay 1 fiducial sequence is compared with
our NGC 416 photometry in
Fig.\ \ref{fig-2}b.
Even though the photometric scatter is large at the main-sequence turnoff
region, one can clearly see that
most of the subgiant branch stars of Lindsay 113 are brighter
than most of the subgiant branch stars of NGC 416 which, in turn,
are brighter than the fiducial subgiant branch of Lindsay 1.
Remembering that these clusters have similar metallicities, we see that
Fig.\ \ref{fig-2} provides graphical evidence suggesting that
the age sequence of these three SMC populous clusters
is Lindsay 1, NGC 416, and Lindsay 113
with Lindsay 1 being the oldest.

Wishing to verify and quantify that this age sequence is correct, we now seek
corroborative evidence in the morphology of the cluster horizontal branches.
Sarajedini \et (\cite{salele1995}; hereafter SLL)
proposed an age indicator for globular clusters with
predominantly red horizontal branches based on the $\bmv$
color difference, $\dbmv$,
between the the mean color of the red horizontal branch (clump)
and the red giant branch at the level of the horizontal branch.
We measured the value of $\dbmv$ in Lindsay 1, NGC 416, and Lindsay 113
following the procedure described by SLL (see Table 1).
The errors in $\dbmv$ are quite small because, by its very nature,
$\dbmv$ is a differential quantity which makes it easy
to measure with high precision.
Table 1 gives our estimates for the cluster
ages and the absolute visual magnitude of the
red horizontal branch clumps, $\mvrhb$, which were
derived using the cluster metallicities and $\dbmv$ values
with Figs.\ 4 and 5 of SLL, respectively.
The reader is reminded that the calibration of
SLL is based on the assumption that the
Galactic globular cluster 47 Tuc is 13 Gyr old and has $\feh = -0.71$ dex.
As expected, the ages inferred from $\dbmv$ have corroborated
the proposed age sequence derived from the relative positions of the
main sequence turnoffs (see Fig.\ \ref{fig-2}):
NGC 416 is younger than Lindsay 1 (by $\sim$2 Gyr)
but older than Lindsay 113 (by $\sim$1.5 Gyr).
The $\dbmv$ ages we find for Lindsay 1 and Lindsay 113 agree well with previous
age estimates in the literature.

\section{DISCUSSION}

The age sequence of the three SMC populous clusters that we have investigated
is Lindsay 1, NGC 416, and Lindsay 113
with Lindsay 1 being the oldest.
{}From the data presented in Table 1 we find that the age {\em{ratios}} of
the younger clusters with respect to Lindsay 1 are
$age_{\rm{NGC416}}/age_{\rm{L1}} \approx 0.73\!\pm\!0.05$
and
$age_{\rm{L113}}/age_{\rm{L1}} \approx 0.52\!\pm\!0.09$
which gives new absolute age estimates of
$age_{\rm{NGC416}} \approx 6.6\!\pm\!0.5$ Gyr
and
$age_{\rm{L113}} \approx 4.7\!\pm\!0.8$ Gyr
assuming the Olszewski \et (\cite{olet1996}) age scale
where the age of Lindsay 1 is 9 Gyr.

Durand \et (\cite{duet1984}) analyzed deep photographic plates
from the 2.5-m du Pont telescope at Las Campanas and produced a
color-magnitude diagram of NGC 416 which they used to estimate an age of
2.5$\pm$0.7 Gyr.
The following year, analysis of integrated photometry produced
significantly older age estimates:
5.9$\pm$1.9 Gyr (Elson \& Fall \cite{elfa1985})
and
8$\pm$1.5 Gyr (Bica \et \cite{biet1985}).
Our estimate of 6.6$\pm$0.5 Gyr is in good agreement with the estimates
based on integrated photometry rather than the estimate based on the analysis
of a color-magnitude diagram.
Why did the supposedly more accurate technique
of CMD analysis produce a significantly lower age estimate than predicted
by integrated photometry?
Durand \et (\cite{duet1984}) found a
concentration of stars in their NGC 416 CMD at
$V \approx 20.5$ mag and $\bmv \approx 0.5$ mag
which they interpreted as being the cluster main-sequence turnoff.
We find no such concentration in our color-magnitude diagrams
and note that the main-sequence turnoff of NGC 416 is below the
apparent plate limit of the Durand \et (\cite{duet1984}) study;
this strongly suggests that the concentration was probably an artifact
of their photometric analysis/subtraction process.

While integrated photometry of NGC 416 did produce a better age estimate
than that based on deep photographic photometry, ages determined from
integrated photometry can also be unreliable.
Elson \& Fall (\cite{elfa1985}) determined that NGC 416 and Lindsay 113
have s-parameter values of 46 and 49, which translated to ages
of 5.9$\pm$1.9 Gyr and 11$\pm$3 Gyr, respectively, using their Eq.\ 1.
They determined that Lindsay 113 was 5 Gyr {\em older} then NGC 416
while we have found that Lindsay 113 is $\sim$30\% ($\sim$1.5 Gyr)
{\em younger} than NGC 416.
Contaminating young
main-sequence
SMC field stars near Lindsay 113 (see our Fig.\ \ref{fig-2}a)
may have contributed enough ultraviolet light to cause the cluster
$(U\!-\!B)$ color within the aperture
to be measured too small (blue) which would have caused the
s-parameter value (hence age) for Lindsay 113 to be overestimated
(cf.~Geisler \et \cite{geet1997}).

Finally, we note that
the age estimates we have given in Table 1 for NGC 416 and Lindsay 113
were derived by extrapolating Fig.\ 4 of SLL to account for the fact that
the measured $\dbmv$ values of these young clusters
are smaller than $\sim$0.15 mag.
Accurate (relative) ages
for the intermediate-age populous
clusters in the Small Magellanic Cloud
(e.g.\ via deep main sequence photometry)
would allow the $\dbmv$ method of SLL
to be {\em{recalibrated}} with
star clusters that are significantly younger than 7 Gyr.
Needless to say, these observations would also provide unique
observational constraints on helium-core burning Population II stellar
evolutionary models.
Such an extended $\dbmv$ method could
prove be a very useful age diagnostic for future studies of
the intermediate-age metal-poor stellar populations in
Local Group galaxies where accurate main-sequence turnoff photometry at
$M_V\!\approx\!+4$ mag is currently not possible or practical.

\acknowledgments

KJM
was supported by a grant from
the National Aeronautics and Space Administration (NASA),
Order No.\ S-67046-F, which was awarded by
the Long-Term Space Astrophysics Program (NRA 95-OSS-16).
AS
was supported by the
NASA
grant number HF-01077.01-94A from
the Space Telescope Science
Institute, which is operated by the Association of Universities for
Research in Astronomy, Inc., under NASA contract NAS5-26555.
AS wishes to thank Lick Observatory for their generosity and
hospitality during his visit.

%
%

\clearpage
\newpage


\def\fig1cap{
\label{fig-1}
The $V$ \vs $\bmv$ color-magnitude diagram of the observed stellar
field in the SMC populous cluster NGC 416.
(a) The 8513 stars with
signal-to-noise ratios S/N$\geq$10 in
both filters are plotted (dots) along with the
2226 CCD/image defects (open circles).
(b) The 3351 stars found on the PC1 CCD.
(c) The 5162 stars found on the WF2, WF3, and WF4 CCDs.
(d) The ``cleaned'' color-magnitude diagram of NGC 416
contains 2826 stars.
The error bars indicate rms ($1\sigma$)
uncertainties for a single star at the corresponding magnitude.
}

\ifundefined{showfigs}{
  \figcaption[mighell.fig1.eps]{\fig1cap}
}\else{
  \begin{figure}[p]
  \figurenum{1}
  \plotone{mighell.fig1.eps}
  \vskip -4cm
  \caption[]{\baselineskip 1.15em \fig1cap}
  \end{figure}
}
\fi


\def\fig2cap{
\label{fig-2}
(a) Our fiducial sequence of Lindsay 1 (Olszewski \et \protect\cite{olet1987})
is compared with our transformed photometry of Lindsay 113 
(Mould \et \protect\cite{moet1984}).
(b) Similarly, our fiducial sequence of Lindsay 1 is compared with our 
photometry of NGC 416. The data for Lindsay 113 and NGC 416 were shifted by
$\Delta V =\,$+0.19 and -0.40 mag, respectively, to match the $V_{\rm{HB}}$ of
Lindsay 1. The adopted reddenings for this figure are taken from Table 1.
The age sequence of these three SMC populous clusters is Lindsay 1, NGC 416, 
and Lindsay 113 with Lindsay 1 being the oldest.
}

\ifundefined{showfigs}{
  \figcaption[mighell.fig2.eps]{\fig2cap}
}\else{
  \clearpage
  \newpage
  \begin{figure}[p]
  \figurenum{2}
  \vskip -6cm
  \plotone{mighell.fig2.eps}
  \vskip -0cm
  \caption[]{\baselineskip 1.15em \fig2cap}
  \end{figure}
}
\fi


\clearpage
\newpage
\begin{deluxetable}{lccccc}
\tablenum{1}
\tablewidth{0pc}
\tablecaption{Cluster Parameters}
\tablehead{
\colhead{Cluster} &
\colhead{[Fe/H]} &
\colhead{E$(\!B\!-\!V\!)$} &
\colhead{$d_{(B\!-\!V)}$} &
\colhead{Age} &
\colhead{$M_V^{\rm{RHB}}$}
\\
\colhead{}&
\colhead{} &
\colhead{(mag)} &
\colhead{(mag)} &
\colhead{(Gyr)} &
\colhead{(mag)}
}
\startdata
 Lindsay 113 & $-1.24$$\pm$$0.11$ & $0.00$$\pm$$0.02$ & $0.117$$\pm$$0.006$ &
$4.0$$\pm$$0.7$ & $0.32$$\pm$$0.04$ \cr
            &                  &                 &                    & $[
4.7$$\pm$$0.8 ]$ \cr
 NGC 416     & $-1.44$$\pm$$0.12$ & $0.08$$\pm$$0.03$ & $0.128$$\pm$$0.004$  &
$5.6$$\pm$$0.3$ & $0.40$$\pm$$0.02$ \cr
            &                  &                 &                    & $[
6.6$$\pm$$0.5 ]$ \cr
 Lindsay 1   & $-1.35$$\pm$$0.08$ & $0.06$$\pm$$0.02$ & $0.152$$\pm$$0.005$  &
$7.7$$\pm$$0.4$ & $0.49$$\pm$$0.02$ \cr
            &                  &                 &                    & $[
\equiv 9.0 ]$ \cr
\enddata
\end{deluxetable}

\end{document}